# Three New Light Curves and Updated Transit Timings of WASP-135 b

**Kristine Kate Torres**
*Rizal Technological University, Boni Avenue, Mandaluyong, 1550 Metro Manila, Philippines; kristinekate.torres@gmail.com*

**Isabela G. Huckabee**
*Department of Astronomy, Cornell University, 122 Sciences Drive, Ithaca, NY 14853, and Ateneo de Manila University, Katipunan Ave, Quezon City, 1108 Metro Manila, Philippines; igh7@cornell.edu*

**Patrizia Phem Odo**
*Rizal Technological University, Boni Avenue, Mandaluyong, 1550 Metro Manila, Philippines; patriziaphem16.odo@gmail.com*

**Danielle Baldono**
*Ateneo de Manila University, Katipunan Avenue, Quezon City, 1108 Metro Manila, Philippines; danielle.baldono@student.ateneo.edu*

**Riana Gabrielle Gamboa**
*University of the Philippines, Roxas Avenue, Diliman, Quezon City, 1101 Metro Manila, Philippines; rianagabriellegamboa@gmail.com*

**Ma. Francelen Rose Molod**
*Ateneo de Manila University, Katipunan Avenue, Quezon City, 1108 Metro Manila, Philippines; francelen.molod@student.ateneo.edu*

**Michelle Dote**
*National University, Coral Way Street, Pasay City, 1300 Metro Manila, Philippines, and University of Santo Tomas, España Blvd, Sampaloc, Manila, 1009 Metro Manila, Philippines; mitchdote@gmail.com*

**Dave Justine Bantilan**
*Cebu Technological University, M.J. Cuenco Avenue, Corner R. Palma St., 6000 Cebu City, Philippines; dave.justinebantz@gmail.com*

**Juan Migelle Ferido**
*De La Salle University, 2401 Taft Avenue, Malate, Manila, 1004 Metro Manila, Philippines; juan_ferido@dlsu.edu.ph*

**Zachary Lacuesta**
*Adamson University, 900 San Marcelino Street, Ermita, Manila, 1000 Metro Manila, Philippines, and University of Santo Tomas, España Blvd, Sampaloc, Manila, 1009 Metro Manila, Philippines; zachlacuesta@gmail.com*

**Mark Erickson Calunod**
*Philippine Normal University, 104 Taft Avenue, Ermita, Manila, 1000 Metro Manila, Philippines; markericksoncalunod100@gmail.com*

**Franchesca Margarette Visitacion**
*University of the Philippines, Roxas Avenue, Diliman, Quezon City, 1101 Metro Manila, Philippines; ffvisitacion@up.edu.ph*

**Jose Alexis Elimanco**
*University of the Philippines, Roxas Avenue, Diliman, Quezon City, 1101 Metro Manila, Philippines; elimanco.ja@gmail.com*

**Disney Gonzales**
*Rizal Technological University, Boni Avenue, Mandaluyong, 1550 Metro Manila, Philippines; ddgonzales@rtu.edu.ph*

**Robert Zellem**
*NASA Goddard Space Flight Center, Greenbelt, MD 20771; robert.t.zellem@nasa.gov*

**Heath Rhodes**
*Jet Propulsion Laboratory, California Institute of Technology, 4800 Oak Grove Drive, Pasadena, CA 91109; heath.a.rhoades@jpl.nasa.gov*

**Kyle Pearson**
*Jet Propulsion Laboratory, California Institute of Technology, 4800 Oak Grove Drive, Pasadena, CA 91109; k.pearson112@gmail.com*

**Abstract**   We present updated transit timing measurements for the hot Jupiter WASP-135 b using three new ground-based transit observations obtained with Leia, a 0.6-meter telescope operated by NASA's Exoplanet Watch at the Table Mountain Facility. These observations, conducted as part of Exoplanet Watch citizen science initiative, were analyzed with the EXOplanet Transit Interpretation Code (EXOTIC) pipeline to generate high-quality light curves and extract precise mid-transit times. By combining our new data with previously published observations, we refined the planet's ephemeris, reducing uncertainties in both the orbital period and mid-transit time. Our final mid-transit value is 2460585.6563426 ± 0.00001908 $BJD_{TDB}$ and the final period value is 1.4013776 ± 0.0000002 days. Our updated timing solution demonstrates a 92% reduction in mid-transit time uncertainty compared to the original discovery paper and improves the precision of transit forecasts through 2030, which is critical to ensure efficient scheduling of future missions, such as ESA's *Ariel*. This work highlights the critical role of ongoing ground-based observations by students and citizen scientists in maintaining accurate ephemerides, which are essential for planning future space-based follow-up with facilities such as the Hubble and James Webb Space Telescopes. The work in this paper was done as part of the SEDS-PH (Students for the Exploration and Development of Space—Philippines) Upskill Groups, which provides opportunities for Filipinos to participate in space-related projects.

# 1. Introduction

Exoplanets are planets outside of our solar system. Over 5,800 exoplanets have been confirmed, and most have been found via the transit method (Christiansen *et al.* 2025). As a planet moves across the disk of its host star relative to the observer, it will block out a portion of the star's light, creating a dip in brightness over time. An abundance of exoplanet detections creates an abundance of objects that need to be monitored. Professional astronomers may not have the time or resources to continuously monitor already confirmed exoplanets (Zellem *et al.* 2020), so this presents a great opportunity for citizen scientist involvement.

Exoplanet Watch[1], a NASA citizen science project, is important in advancing exoplanet research by enabling public participation in the observation of exoplanets. Citizen scientists contribute to professional-level research by using small, ground-based telescopes to refine the timing of exoplanet transits and improve the precision of ephemerides. As compared to space-based data alone, the collaboration of citizen scientists with professionals is significant to collectively realize exoplanet ephemeris maintenance and obtain accurate predictions of future transit events (Zellem *et al.* 2020; Pearson *et al.* 2022; Noguer *et al.* 2024).

By integrating data from citizen science projects such as Exoplanet Watch, researchers can achieve better constraints on exoplanet ephemerides compared to space telescope data alone. This expanded observational baseline improves transit time predictions, which are critical for maximizing the efficiency and scientific output of expensive, highly competitive space telescopes like the James Webb Space Telescope (JWST) and Hubble Space Telescope (HST).

Over time, uncertainties in a planet's transit timing can accumulate and consequently affect the accuracy of future observations. By continuously updating ephemerides, Exoplanet Watch helps mitigate these uncertainties and improve the scheduling of both ground- and space-based follow-up observations. Thus, the contributions of amateur astronomers and citizen scientists ensure that the future use of expensive time on telescopes is as efficient as possible (e.g. Hewitt *et al.* (2023). As part of the Exoplanet Watch community, we worked to analyze three new transits of WASP-135 b with Leia, the 0.6-meter telescope operated on Table Mountain in California, USA.

WASP-135 b is a transiting hot Jupiter exoplanet discovered through the Wide Angle Search for Planets (WASP) survey. It orbits a G-type main-sequence star (G5V), similar to the Sun, and which has a stellar mass and radius of $0.98 \pm 0.06\ M_\odot$ and $0.96 \pm 0.05\ R_\odot$, respectively. WASP-135 b has a planetary mass of $1.90 \pm 0.08\ M_{Jup}$, a planetary radius of $1.30 \pm 0.09\ R_{Jup}$, and a period of 1.401 days. Its high levels of insolation may contribute to its inflated radius, a phenomenon observed in many close-orbiting gas giants, possibly due to mechanisms like Ohmic heating or kinetic heating that transport stellar energy into the planetary interior (Spake *et al.* 2016).

Hot Jupiters like WASP-135 b provide unique opportunities for studying exoplanet atmospheres and formation mechanisms. Despite extensive studies, the reasons behind their inflated radii and formation so close to host stars remain debated. Also, the age estimates for their host stars present potential evidence for angular momentum transfer, or tidal "spin-up," between these hot Jupiters and their stars. The case of WASP-135 b showed weak evidence of angular momentum transfer, indicating that its host is possibly a young star. Understanding the complex interactions between hot Jupiters and their host stars, including spin-up effects, remains an active area of research, highlighting the importance of further observational data on systems like WASP-135 b (Spake *et al.* 2016).

The following work is done in partnership with the Students for the Exploration and Development of Space—Philippines (SEDS-PH) as part of their Upskill program to develop the skills of Filipino students and professionals interested in space and astrophysics.

# 2. Data

Light curves for the exoplanet WASP-135 b were generated from observational transit data obtained on 09 May 2024, 16 June 2024, and 03 July 2024, using a 0.6-meter Ritchey-Chrétien reflector telescope, located at a latitude of +34.382, longitude of −117.6818, and an elevation of 2286 meters. The first two nights captured full transits, while the third captured a partial transit. All observations used a clear filter and 60s exposure times. The data was collected remotely from the Table Mountain Facility (TMF), which is located near Wrightwood, California. The Table Mountain Facility supports various scientific activities, such as astronomy and atmospheric studies. TMF's 0.6-meter, dubbed Leia, outputs images $2048 \times 2048$ pixels in size and a pixel scale of 0.3 arcseconds per pixel. We show an example image of the data in Figure 1. Leia is mounted on an off-axis German equatorial mount, providing stabilization and precise tracking of celestial objects as the Earth rotates.

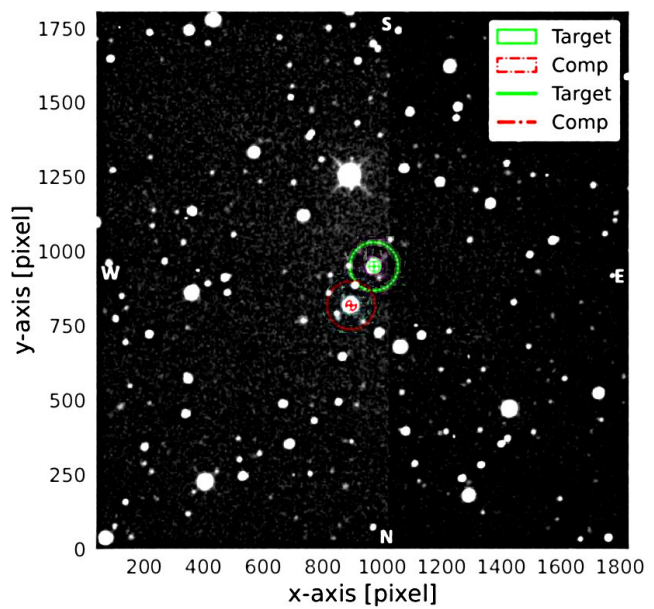

Figure 1. Field of view for WASP-135 b (image scale in arcsec/pix: 0.299772). Example image from 09 May 2024, with the target star, WASP-135, and the comparison star labeled in green and red, respectively.

---

[1] https://exoplanets.nasa.gov/exoplanet-watch/

## 3. Methods overview

EXOTIC (EXOplanet Transit Interpretation Code; NASA (2025)) is the primary software involved in data reduction and visualization for Exoplanet Watch (EpW), a NASA Citizen Science Project aimed at utilizing non-professional, amateur, and student efforts in the production of exoplanet transit data for publications. EXOTIC, although compatible for use in Python3 on local machines, is also designed to be run through Google Colab, a free-to-use Google service. Colab as an online notebook editor finds its efficiency in its simple interface, straightforward approach, and a lack of need for installation, all of which align with citizen science's objective of non-professional encouragement and involvement. Given a user has internet connection, exoplanet light curve data can be reduced by EXOTIC with as little as a smartphone.

The workflow begins with assembling the necessary observational inputs: raw image data in FITS or FITS.gz format, information on target and comparison stars (coordinates and magnitudes), and a corresponding American Association of Variable Star Observers (AAVSO) starfield chart. Once uploaded, EXOTIC integrates these inputs to produce calibrated light curves and associated outputs. The software also generates files formatted for AAVSO submission, along with optional products such as a certificate of participation.

The reduction routine can be divided into six stages, namely tracking the star, extracting the flux of the target, normalization by a comparison star, determining the optimal aperture and comparison star, the full light curve fitting routine, and results interpretation, in that order.

Star tracking is achieved by applying a two-dimensional Gaussian centroiding routine, which corrects for frame-to-frame positional shifts. Flux measurements are then obtained using circular apertures, with background levels subtracted via an annulus surrounding the target. For each transit, the optimal annulus and aperture radius are listed in Table 1. Normalization is performed through differential photometry, using a carefully chosen non-variable comparison star of similar brightness and color.

The target star's signal is determined by its optimal aperture, which balances the image, reducing the background noises to its nearby sources. This method involves choosing an aperture size that covers the star's point spread function. We use the aperture with the lowest noise yield and largest signal-to-noise ratio (SNR). Selecting a comparison star is essential for differential photometry. The comparison star should be non-variable and have a brightness and color similar to that of the target star to avoid errors. However, the AAVSO Variable Star Plotter (VSP) has identified only one non-variable star, which appears near the edge of the field of view (FOV). To compensate for this, the selected comparison stars used for the WASP-135 b reductions are those near and as bright as the target star, as shown in Figure 2 and summarized in Table 2. By doing so, it ensures that the best comparison stars have relatively the same sky conditions as the target star. Also, the use of multiple comparison stars may reduce the systematic errors by averaging out the variations.

After developing the light curve, EXOTIC proceeds to model fitting. EXOTIC fits the time series data and an airmass correction function using a Markov Chain Monte Carlo within the Ultranest sampler (Buchner 2021), with the airmass correction function in the form of:

$$F_{obs} = a_0 e^{a_1\beta} F_{transit} \qquad (1)$$

where $a_i$ are the airmass correction coefficients, $\beta$ is the airmass value, $F_{obs}$ is the detector recorded flux, and $F_{transit}$ is the actual transit signal.

Parameters fitted for using nested sampling include the planet-to-star radius ratio ($R_p/R_s$), orbital inclination, transit duration, and mid-transit time. EXOTIC uses the Bayesian Information Criterion (BIC) values to then perform model selection (Schwarz (1978), where BIC = $\chi^2$ + k ln(n), k being the number of free parameters and n the number of data points).

Table 1. Optimal Annulus and Aperture per Transit.

| Date | Apertur (pixels) | Annulus (pixels) |
|---|---|---|
| 09 May 2024 | 16.45 | 63.42 |
| 16 June 2024 | 0 | N/A |
| 03 July 2024 | 5.22 | 28.19 |

Table 2. Coordinates of the target and comparison stars and comparison star usage.

| Star | R.A. h m s | Dec. ° ' " | Date used |
|---|---|---|---|
| WASP-135 (Target) | 17 49 08.39 | +29 52 44.7 | |
| Comparison Star #1 | 17 49 06.627 | +29 53 23.03 | |
| Comparison Star #2 | 17 49 02.981 | +29 51 53.10 | 09 May 2024 |
| Comparison Star #3 | 17 48 59.167 | +29 50 49.04 | 16 June, 03 July 2024 |

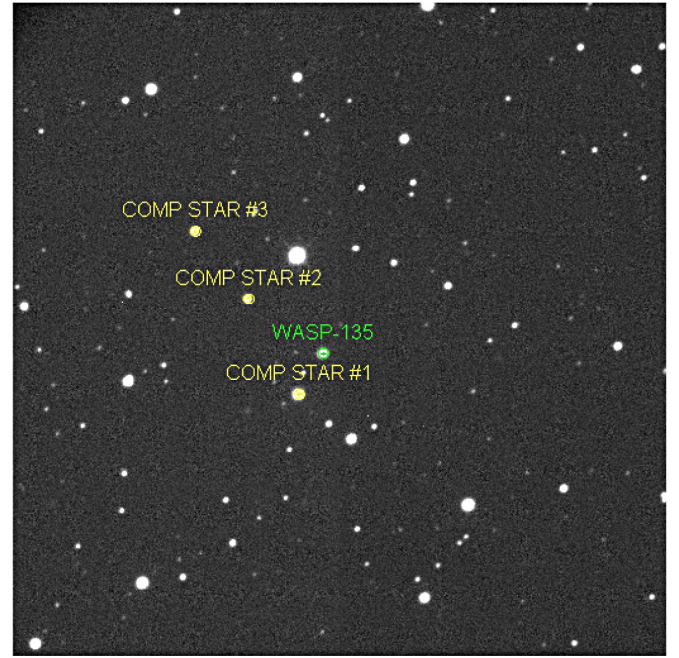

Figure 2. Annotated star field of WASP-135 using AstroImageJ (Collins *et al.* 2017). The target star is annotated with a green crosshair, while the comparison stars are marked in yellow.

significance of 3σ. This threshold provides strong evidence against the null hypothesis of no transit signal.

$$\frac{\text{Transit Depth}}{\text{Transit Depth Uncertainty}} \geq 3 \quad (2)$$

We then compared the mid-transit times of our observations with mid-transit times from the NASA Exoplanet Archive, TESS, and Exoplanet Watch in the observed-calculated (O–C) plot shown in Figure 4. The NASA Exoplanet Archive mid-transit times are from Spake *et al.* (2016), Kokori *et al.* (2023), and Ivshina and Winn (2022). σ-clipping is utilized to remove outliers from the total O–C dataset. To do this, we loop through and compute the median residual value, the median uncertainty on each data point, the standard deviation, and the combined uncertainty, defined by the square root of the sum of the squared quantities of the standard deviation and the median uncertainty of the dataset. Afterward, we create an array of the residuals which is then normalized by the combined uncertainty and its uncertainty. Further, the elements of the array that exceed the 3σ range are identified and removed from the dataset. Afterward, the program recomputes for the median, median uncertainty, standard deviation, and combined uncertainty and it loops until there are no more elements of the array that exceed the recomputed 3σ range. We found that were no outliers outside of the 3σ range and therefore no data points were removed from our analysis.

Using our ephemeris fitter, we calculated the mid-transit time to be $T_{mid}$ = 2460585.6563426 ± 0.0001908 $BJD_{TDB}$ and period P = 1.4013776 ± 0.0000002 days. We summarize this result in Table 3, along with previously published values.

The data significantly reduced the uncertainty in the timing of the planet's transit compared to earlier research. The updated mid-transit time is sparingly more precise than previous estimates. Compared to the discovery paper of Spake *et al.* (2016), the data show reduced uncertainty of both mid-transit time and period. Moreover, our results are consistent with the recent study of Kokori *et al.* (2023), where our calculated uncertainty value of 0.0001908 BJD_TDB is close to their value of 0.00019 BJD_TDB. Additionally, we reduced the uncertainty in our calculated orbital period compared to previous studies by Spake *et al.* (2016), Kokori *et al.* (2023), and Ivshina and Winn (2022).

As mid-transit time uncertainties become larger over each orbit, we forward propagated both our mid-transit time as well as the previously published values. To achieve this, we used a modified equation from Zellem *et al.* (2020):

$$T_{mid} = n_{orbit}^2 \times \Delta P + T_0$$
$$\Delta T_{mid} = \sqrt{(n_{orbit}) \times (\Delta P)^2 + (\Delta T_0)^2} \quad (3)$$

and calculated our mid-transit time forward propagated to 2030 to be 2462502.7408994 ± 0.0003336 $BJD_{TDB}$.

Compared to the transit times reported in the discovery paper by Spake *et al.* (2016), our uncertainty in mid-transit time was reduced by 92.1%. Similarly, our uncertainty is reduced

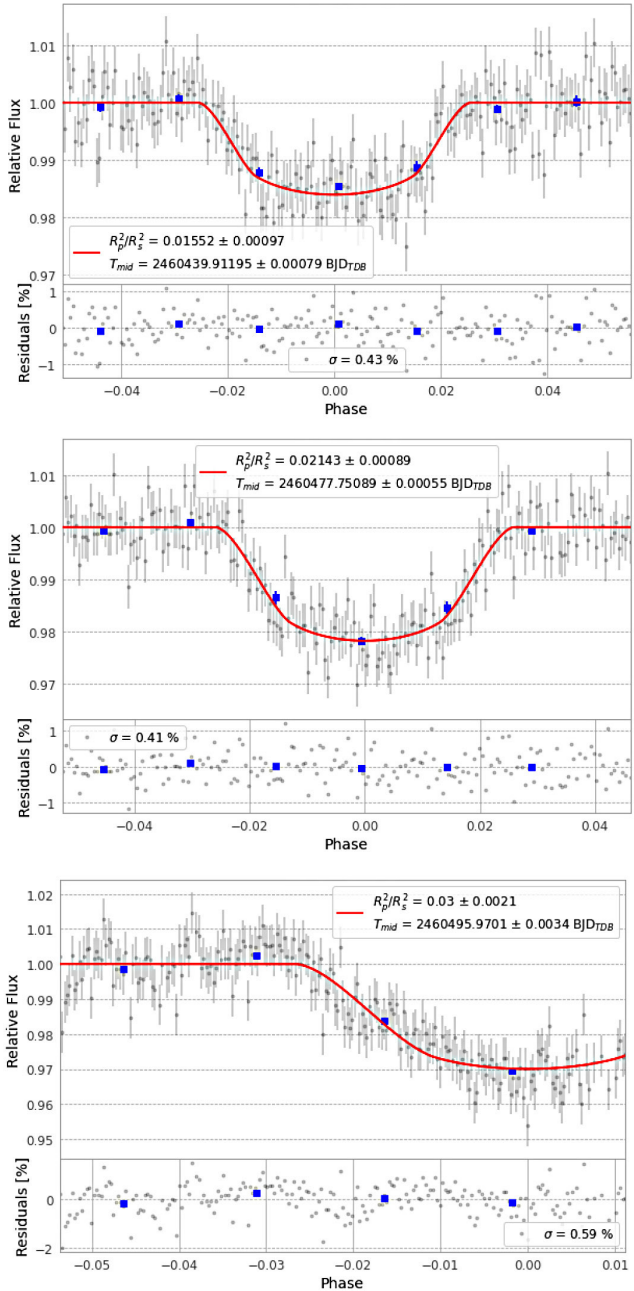

Figure 3. Light curves of the WASP-135 b transit processed with EXOTIC from 09 May 2024 (top), 16 June 2024 (middle), and 03 July 2024 (bottom).

Airmass and limb-darkening constraints are also accounted for in fitting the light curves.

## 4. Results

To ensure consistency, the datasets were reduced independently by multiple people on our team. We then kept the results with the lowest residuals, as shown in Figure 3. In the reduction process, EXOTIC used the same comparison star for all three observations.

In addition, we confirmed all observations to have statistically significant transit detections. We considered a detection significant if the signal-to-noise ratio (transit depth divided by its uncertainty) exceeded 3, corresponding to a detection

Table 3. Mid-transit times and orbital periods from various sources.

| $T_{mid}$ (BJD$_{TDB}$) | $\Delta T_{mid}$ (d) | $P$ (d) | $\Delta P$ (d) | Source |
|---|---|---|---|---|
| 2460585.65634 | 0.00019 | 1.4013776 | 0.00000020 | Our data |
| 2455230.9902 | 0.0009 | 1.4013794 | 0.0000008 | Spake *et al.* (2016) |
| 2459046.94408 | 0.00019 | 1.40137864 | 0.00000039 | Kokori *et al.* (2023) |
| 2459010.50778 | 0.00025 | 1.40137841 | 0.00000034 | Ivshina and Winn (2022) |

*Note: The quoted uncertainties are the 1σ uncertainties.*

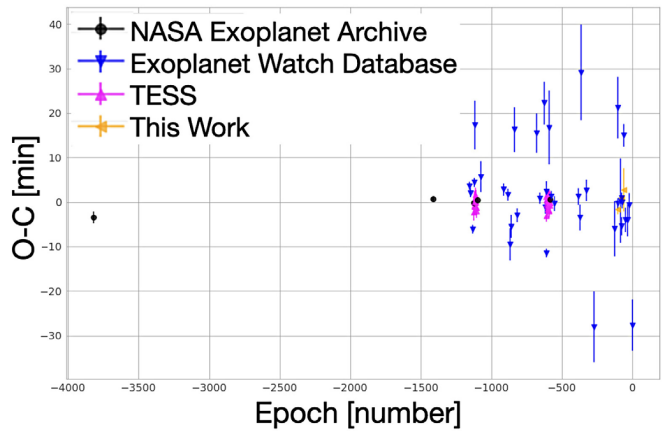

Figure 4. Observed – Calculated (O–C) plot for WASP-135 b. This plot displays the deviations of the actual mid-transit times we observe from what we calculated them to be given a circular, two-body system. Mid-transit times from the NASA Exoplanet Archive are shown in black circles, TESS in magenta triangles, Exoplanet Watch in blue upside-down triangles, and our observations from TMF in orange left-pointing triangles. Most of the data spread is from Exoplanet Watch, which is expected given these telescopes are generally smaller and ground-based.

by 66% compared to Kokori *et al.* (2023). After forward-propagating the mid-transit times published by Spake *et al.* (2016) and Kokori *et al.* (2023) to 2030, the new mid-transit times were found to be 2462502.7479 ± 0.0042 BJD$_{TDB}$, and 2462502.74381 ± 0.00098 BJD$_{TDB}$, respectively.

### 5. Conclusion

We present three new transit light curves of the Hot Jupiter WASP-135 b obtained with Leia, the 0.6-m telescope operated by NASA's Exoplanet Watch on Table Mountain. We update the mid-transit time and orbital period for WASP-135 b and forward propagate this mid-transit time to 2030. In doing so, we lower the uncertainties on both the mid-transit time and orbital period compared to previously published values and demonstrate how ground-based efforts by students and citizen scientists play an important role in keeping these exoplanet transit times up-to-date for use in future in studies.

### 6. Acknowledgements

This publication makes use of data products from Exoplanet Watch, a citizen science project managed by NASA's Jet Propulsion Laboratory on behalf of NASA's Universe of Learning. This work is supported by NASA under award number NNX16AC65A to the Space Telescope Science Institute, in partnership with Caltech/IPAC, Center for Astrophysics|Harvard & Smithsonian, and NASA Jet Propulsion Laboratory.

This research has made use of the NASA Exoplanet Archive, which is operated by the California Institute of Technology, under contract with the National Aeronautics and Space Administration under the Exoplanet Exploration Program. Data in this research were collected at the Jet Propulsion Laboratory Table Mountain Facility owned by the Jet Propulsion Laboratory, California Institute of Technology, under contract with the National Aeronautics and Space Administration.